\documentclass[english,twocolumn,aps,prl,showpacs,superscriptaddress]{revtex4}
\usepackage[T1]{fontenc}
\usepackage[latin9]{inputenc}
\usepackage{array}
\usepackage{booktabs}
\usepackage{graphicx}



\usepackage{babel}

\begin{document}

\title{Resonant phonon coupling across the La$_{1-x}$Sr$_{x}$MnO$_{3}$/SrTiO$_{3}$
interface}

\author{Y. Segal}

\affiliation{Department of Applied Physics, Yale University, New Haven CT 06520-8284}

\affiliation{Center for Interface Structure and Phenomena, Yale University, New Haven CT 06520-8284}

\author{K. F. Garrity}

\affiliation{Center for Interface Structure and Phenomena, Yale University, New Haven CT 06520-8284}

\affiliation{Department of Physics, Yale University, New Haven CT 06520-8120}

\author{C. A. F. Vaz}

\affiliation{Department of Applied Physics, Yale University, New Haven CT 06520-8284}

\affiliation{Center for Interface Structure and Phenomena, Yale University, New Haven CT 06520-8284}

\author{J. D. Hoffman}

\affiliation{Department of Applied Physics, Yale University, New Haven CT 06520-8284}

\affiliation{Center for Interface Structure and Phenomena, Yale University, New Haven CT 06520-8284}

\author{F. J. Walker}

\affiliation{Department of Applied Physics, Yale University, New Haven CT 06520-8284}

\affiliation{Center for Interface Structure and Phenomena, Yale University, New Haven CT 06520-8284}

\author{S. Ismail-Beigi}

\affiliation{Department of Applied Physics, Yale University, New Haven CT 06520-8284}

\affiliation{Center for Interface Structure and Phenomena, Yale University, New Haven CT 06520-8284}

\affiliation{Department of Physics, Yale University, New Haven CT 06520-8120}

\author{C. H. Ahn}

\affiliation{Department of Applied Physics, Yale University, New Haven CT 06520-8284}

\affiliation{Center for Interface Structure and Phenomena, Yale University, New Haven CT 06520-8284}

\affiliation{Department of Physics, Yale University, New Haven CT 06520-8120}

\begin{abstract}
The transport and magnetic properties of correlated La$_{0.53}$Sr$_{0.47}$MnO$_{3}$ ultrathin films, grown epitaxially on SrTiO$_{3}$,
show a sharp cusp at the structural transition temperature of the substrate.
Using a combination of experiment and theory we show that the cusp
is a result of resonant coupling between the charge carriers in the film and a soft phonon mode
in the SrTiO$_{3}$, mediated through oxygen octahedra in the film. The amplitude of the mode diverges towards the transition temperature, and phonons are launched into the first few atomic layers of the film affecting its electronic state.
\end{abstract}

\pacs{64.70.K-;68.35.Ja;73.50.-h;75.70.Ak;}

\maketitle
The coupling of phonons to charge carriers is a process of key importance
for a broad set of phenomena, ranging from carrier mobility in semiconductors
to Cooper pairing. In recent times, phonon effects at interfaces emerged as a 
topic of great importance in the understanding and
design of nano-structured materials \citep{interfacialphonons}. Coupling between charge, structure and magnetic ordering
is particularly strong in the Mn oxides \citep{RefWorks:462}, which are used as a component in heterostructure multiferroics \citep{CarlosPRLPaper}. In
these materials, localized spins and mobile carriers reside on the
Mn sites, each surrounded by an oxygen octahedra. Intersite hopping
occurs through orbital overlap of the Mn with neighbouring oxygens,
making it highly sensitive to the static orientation of the octahedra
and to phonons that alter the octahedra's orientation
\citep{RefWorks:463}. This interplay between structure and properties
has been exploited to control the electronic phase of CMR films via
strain, and also via coherent photoexcitation of a specific octahedra
vibration mode \citep{RefWorks:454}. \\
In this Letter, we use a specially designed thin film device to
isolate and characterize phonon-carrier coupling within a few atomic
layers of an interface between the perovskite SrTiO$_{3}$ (STO) and
the CMR oxide La$_{0.53}$Sr$_{0.47}$MnO$_{3}$ (LSMO). A soft octahedral
rotation phonon with a divergent amplitude in the STO couples to the
corresponding mode of the film. This coupling results in a marked change in
the electronic and magnetic properties, including a sharp cusp in
the resistivity and a dip in the magnetic moment. The sensitivity
of LSMO to octahedra orientation allows us to experimentally probe the microscopic
character of this interfacial phonon coupling, and compare it to theory.
The thin film devices consist of La$_{0.53}$Sr$_{0.47}$MnO$_{3}$
films grown by molecular beam epitaxy on TiO$_{2}$-terminated STO
(001) substrates and overlaid by Pb(Zr$_{0.2}$Ti$_{0.8}$)O$_{3}$
(PZT), which is used to provide ferroelectric field effect modulation of the number and distribution
of carriers in the film. Details concerning fabrication and structural
characterization are described elsewhere \citep{CarlosGrowthPaper}. In the
bulk LSMO phase diagram, the $x=0.5$ composition separates the ferromagnetic
metallic phase from an insulating antiferromagnetic phase \citep{RefWorks:476}.
When grown commensurate to the STO, the substrate induces tensile
strain in the film, which is known to stabilize an A-type antiferromagnetic
metallic phase (AF-M) \citep{RefWorks:414}. Using X-ray diffraction,
we verified that our films are under tensile strain, with $c/a=0.975$,
in agreement with previous studies \citep{RefWorks:414}.

\begin{figure}
\includegraphics[clip,width=8.5cm]{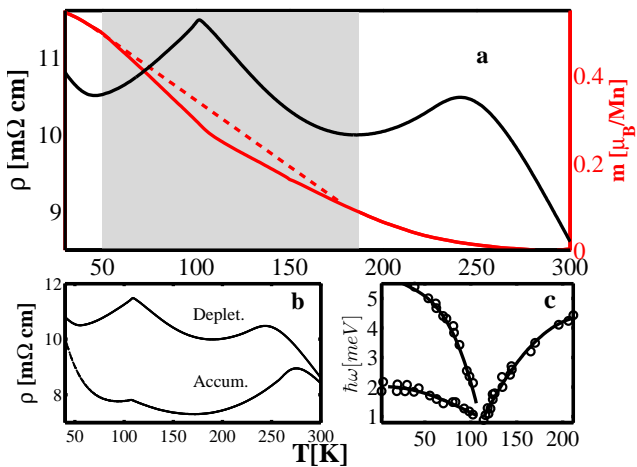}\caption{\label{fig:Transport}Enhanced carrier-phonon scattering. (a) Left
axes: Resistivity of an 11\,uc La$_{0.53}$Sr$_{0.47}$MnO$_{3}$
film showing a strong cusp at 108\,K. The PZT overlayer is in the
depletion state. Right axis: Magnetic moment of a 15\,uc La$_{0.55}$Sr$_{0.45}$MnO$_{3}$
film. The moment is measured along the {[}100{]} direction under an
applied magnetic field of 1kOe. A dip in the moment is observed, overlapping
the temperature range of the resistivity cusp (emphasized by grey
box). The dashed line is a linear interpolation between the edges
of the dip region. b) The resistivity of the 11\,uc film
for the two polarization states of the PZT. c) Energy of the $\Gamma_{25}$ phonon mode in STO, showing the
softening around the STO transition temperature (after Ref.\,\citealp{RefWorks:458}). 
Lines are a guide to the eye. Below the structural phase transition the mode splits due to the breaking of cubic symmetry.}\end{figure}

Transport measurements of an 11\,unit cell (uc) LSMO film are shown
in Fig.\,\ref{fig:Transport}a. The broad peak in resistivity at
250\,K corresponds to a metal-insulator transition, typical of this
material. In addition, a unique feature is observed in our films:
a large and sharp resistance peak centered around 108\,K, which corresponds
to the temperature of the STO soft phonon peak. We observe further
that the magnitude of the resistivity cusp decreases when the thickness of the film increases by a few unit cells.
Indeed, in previous studies of films $\approx$80\,uc thick, only a trace of this feature was
observed \citep{RefWorks:477}. This film thickness dependence implies
that the strength of the mechanism creating the cusp decays quickly
away from the STO/LSMO interface. We can verify this by switching
the polarization state of the PZT. When the PZT is switched to the
``depletion'' state, holes are removed from the top layer of the LSMO,
pushing the conducting region closer to the substrate. The opposite
occurs in the ``accumulation'' state \citep{CarlosPRLPaper}. We find
that the PZT has a pronounced effect on the cusp (Fig.\,\ref{fig:Transport}b),
making it much larger in the depletion state, in agreement with the
notion of a rapid decay into the film. We note, however, that presence
of PZT is not required to observe the effect:
the same features are observed on uncapped LSMO films. We also observe a striking dip in the magnetic moment centered around the STO transition
temperature (Fig.\,\ref{fig:Transport}a). While the majority of
the LSMO is in an antiferromagnetic-metallic state, a small ferromagnetic
component remains \citep{RefWorks:477}. The dip in magnetic moment
corresponds to a decrease in magnetic order within the ferromagnetic
phase.\\

We attribute the transport and magnetism anomaly to a coupling between
the LSMO and the phonon softening that occurs in STO around the 108\,K
structural transition. The $\Gamma_{25}$ $(111)$ zone edge phonon
\citep{RefWorks:460,RefWorks:458} becomes lower in energy as the
transition is approached from both temperature directions. Fig.\,\ref{fig:Transport}c,
reproduced from Ref.\,\citealp{RefWorks:458}, shows the $\Gamma_{25}$
phonon energy as a function of temperature. The softening leads to
a divergent increase in mode occupation or amplitude. The motion
associated with this mode is a rotation of the TiO$_{6}$
octahedra. Below the transition temperature, the octahedra stabilize into a rotated antiferrodistortive
(AFD) configuration accompanied by a tetragonal distortion of the
unit cell. Since the film is mechanically constrained to the substrate
at the atomic level, motions of the TiO$_{6}$ octahedra couple to the MnO$_{6}$ ones, inducing both static and dynamic
changes in their configuration. 

\begin{figure}
\includegraphics[clip,scale=0.3]{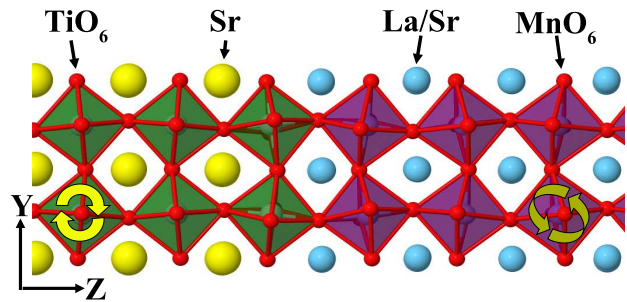}
\caption{Side view of STO-LSMO interface geometry. The plot shows calculated ground-state atomic positions. Away from
from the interface, the STO is fixed to have bulk-like octahedral rotations around the $x$ axis (into the page). The LSMO
geometry at the interface is modified by the STO; however, the LSMO relaxes to its bulk-like octahedral rotations around both in-plane axes within 2-3 unit cells.
\label{fig:interface}}\end{figure}

We examine two mechanisms whereby the resistance of the LSMO layer might increase: $(i)$ static changes of the LSMO 
structure causing a change of electronic band parameters; $(ii)$
decreased carrier relaxation times due to enhanced phonon scattering,
i.e. a dynamic effect. The static and dynamic contributions are reflected
in the expression for the conductivity in the relaxation time approximation
$\sigma_{ij}\propto\tau m_{ij}^{-1}$, where $\tau$ is the relaxation
time and $m_{ij}^{-1}$ is the reciprocal effective mass tensor \citep{ashcroft}. 

To treat the temperature-dependent character of the coupling phenomena,
we perform finite temperature simulations by building a classical
 model of the energetics of the system as a function of oxygen
displacements. Our model includes harmonic coupling between oxygens,
4$^{th}$ order on-site anharmonic terms to stabilize the symmetry
breaking, and lowest order coupling between oxygen displacements and
stress, thus capturing the STO phase transition \citep{sto1}. Model
parameters are obtained via density functional theory calculations
using the spin-polarized PBE GGA functional \citep{GGA} and ultrasoft
pseudopotentials \citep{ultrasoft}. Ground states for both bulk strained
LSMO (using the virtual crystal approximation \citep{vca_vand}) and
the LSMO/STO system are calculated, reproducing the experimental A-type
ordering. In addition, the ground state calculation shows how the octahedral orientation is continuous going from substrate to film (see Fig.\,\ref{fig:interface}), as was recently shown in a similar theoretical study \cite{rondinellioctahedra}. The harmonic interatomic force constants are calculated
with DFT perturbation theory (DFPT) \citep{ModelH2}, and the remaining
parameters are fit to strained bulk calculations. We then perform
classical Monte Carlo sampling on this model in a periodic box. The
box contains $10\times10\times100$ perovskite unit cells composed
of 60 STO and 40 LSMO unit cells in the $z$ direction. 

To evaluate the role of static structural changes, we compute the
conductivity tensor of bulk strained LSMO for the static octahedra
rotation angles obtained from the Monte-Carlo model. The conductivity
is calculated from direct first principles evaluation of the reciprocal
effective mass tensor, by summing over all bands at the Fermi energy
\citep{ashcroft}. The upper bound of conductivity change is estimated
by using the angles at the LSMO/STO interface, which change the most
due to the substrate-film coupling. We find that the static coupling effect
appears only below the phase transition temperature. The phase transition
causes the octahedra angles in the LSMO to increase somewhat; however,
the magnitude of resulting change in conductivity is too small to
account the experimental findings. The computed Mn-O-Mn hopping elements change only by 1-2\% due to
static structural changes, while the experimental conductivity changed by more than 10\%. 

\begin{figure}
\includegraphics[clip,scale=0.2]{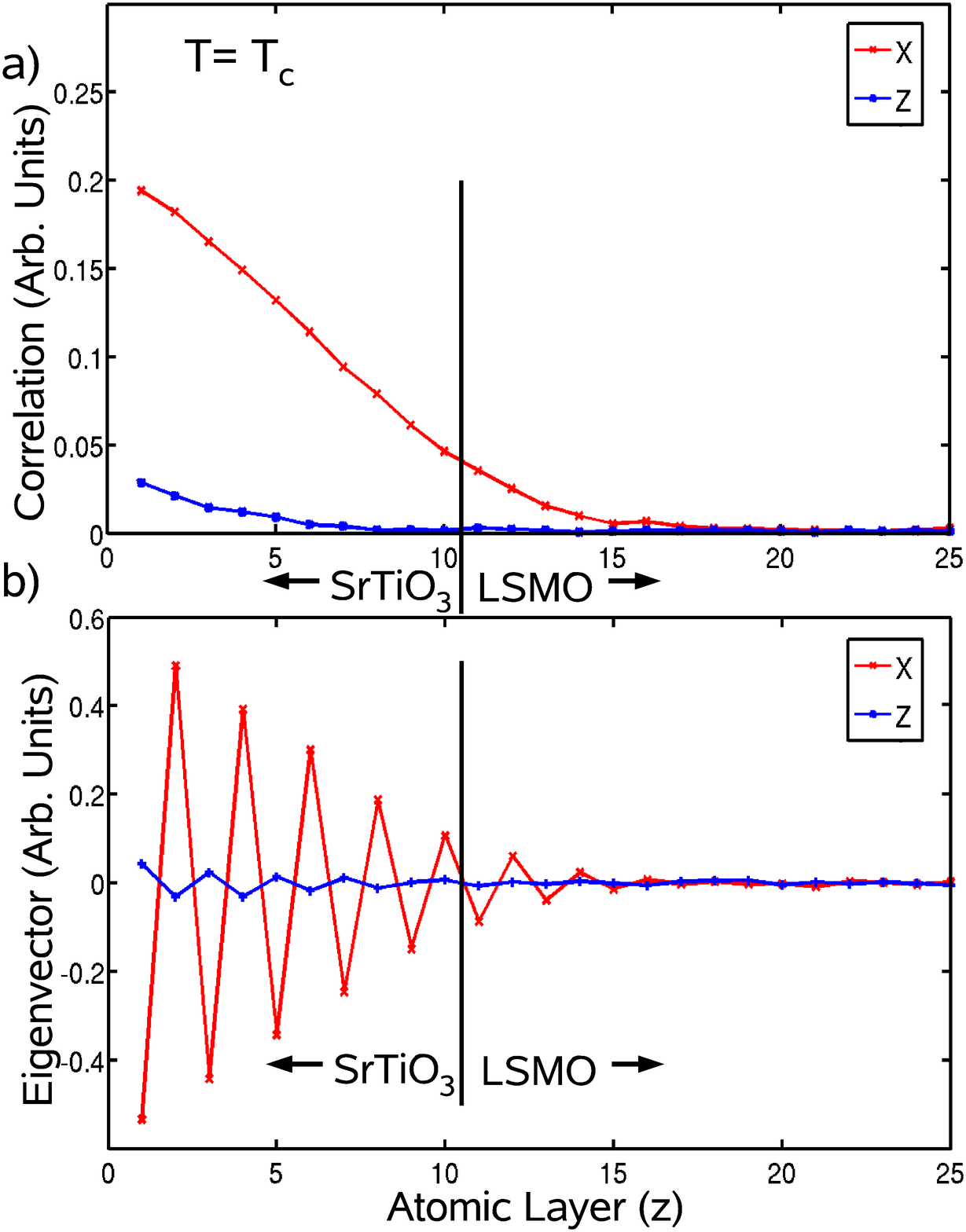}

\caption{\label{fig:correlation}DFT/Monte Carlo results of phonon transfer
from STO to LSMO. a) Absolute magnitude of the correlation function
of oxygen displacements with those deep in STO versus layer number
at $T=T_{C}$. $x$ and $z$ indicate displacements stemming from
octahedra rotation around the $x$ and $z$ axis respectively. (The
$y$ component is equal to $x$ by symmetry). b) $x$
and $z$ components of the lowest frequency eigenvector at $T=T_{C}$.
Layer-to-layer sign changes reflect the AFD nature of the oxygen displacments. }

\end{figure}

Because the static structural change manifests only below 108\,K
and does not yield a large enough change in conductivity, we examine
whether the $\Gamma_{25}$ phonon might extend into the LSMO and cause
dynamic carrier scattering. Hence, we compute the oxygen-oxygen correlation
matrix $c_{ij}=\langle x_{i}x_{j}\rangle-\langle x_{i}\rangle\langle x_{j}\rangle$
where $x_{i}$ are the oxygen displacements from their equilibrium
pseudocubic positions. We find that near the STO phase transition,
the correlation length in STO diverges, as expected. Furthermore,
oxygen motions in the interfacial LSMO layers become correlated with
those deep in the STO (Fig.\,\ref{fig:correlation}a) demonstrating
that STO soft phonons extend into the LSMO. We quantify this relation
more precisely by extracting the dominant eigenvectors of the correlation
matrix, which are the softest phonon modes in the finite-temperature
harmonic system. Fig. \ref{fig:correlation}b shows the lowest frequency
eigenvector: it decays exponentially into the LSMO with a decay
length of 2.3 unit cells.

\begin{figure}
\includegraphics[clip,scale=0.4]{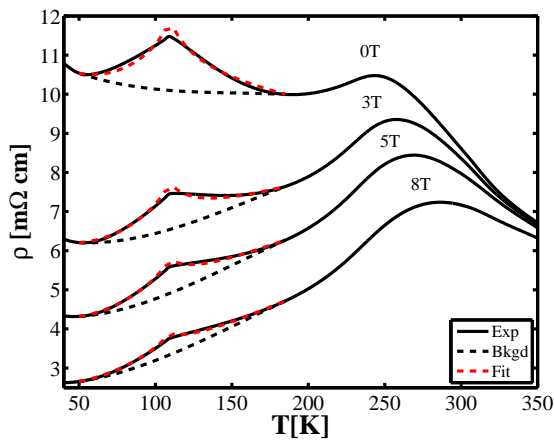}\caption{\label{fig:fits}Model fitting: resistivity of an 11\,uc film of
La$_{0.53}$Sr$_{0.47}$MnO$_{3}$ as a function of temperature, for
several out-of-plane magnetic field values. The PZT overlayer is in
the depletion state. Black dashed lines are interpolations excluding
the cusp and red dashed lines are fits to $\Gamma_{25}$ phonon coupling.}

\end{figure}

Building upon our theoretical results, we use the following simple
model to fit the experimental resistivity data: the scattering due
to the soft mode is $cne^{-2z/\lambda}$, where $n$ is the $\Gamma_{25}$
occupation number in STO (given by the Bose-Einstein function using
the energies in Fig.\,\ref{fig:Transport}c). $c$ is a conversion
factor from $n$ to resistivity (linearly related in phonon scattering
theory \citep{Ziman}); and $\lambda$ is the decay length of the
induced octahedra motion amplitude ($\lambda/2$ is the decay length
for the phonon number). The total conductivity is obtained by summing
over the film layers: $\sigma=\sum\limits _{layers}(\rho_{\mathrm{base}}+cne^{-2z/\lambda})^{-1}$.
$\rho_{\mathrm{base}}$ is the unperturbed LSMO resistivity, which
we find by passing a smooth line under the cusp, following the method
used in an electron paramagnetic resonance study of the softening-induced
disorder in STO \citep{PhysRevB.7.1052}. This model describes the
experimental data well, as shown in Fig.\,\ref{fig:fits}, and yields
a decay length of 1.8\,uc, in good agreement with theory. This verifies
our attribution of the resistivity cusp to resonant coupling of
the divergent $\Gamma_{25}$ mode into the LSMO, where it strongly
scatters the carriers by disturbing the Mn-O-Mn hopping path. 

The phonon coupling picture also explains the dip in magnetic moment,
when phonon-magnon interactions are taken into account. For the manganites,
it has been shown theoretically \citep{RefWorks:484} that when phonons
involving Mn or O distortions are added to the Heisenberg Hamiltonian,
the magnon spectrum is softened. The phonons injected from the substrate
cause a softening of the magnon spectrum in the LSMO, with maximal softening occuring at the STO transition. Magnon occupation
increases with spectrum softening, leading to the reduction in the
magnetic moment. The overlap between the temperature range of the
transport cusp and the moment dip is striking (Fig.\,\ref{fig:Transport}a),
confirming that they are both driven by $\Gamma_{25}$ phonon softening.\\
Previous work on the influence of the STO transition on manganite
films dealt with effects attributed to the appearance of $a$ and
$c$ domains in the STO and the resulting change in the strain state
of the film below the transition temperature \citep{RefWorks:432,RefWorks:434,RefWorks:431}.
In the current work, we find that effects appear both above and below
the transition temperature and correspond to the temperature range
of phonon softening. In addition, the short decay length that we
find does not agree with a strain-mediated phenomenon. The LSMO remains
strained to the substrate up to a thickness of at least 80 uc \citep{RefWorks:414},
so that changes related to strain should be evident at these thicknesses
as well. These two facts preclude the strain configuration of the
STO below the transition from being the source of the cusp. To further
verify that the cusp feature is independent of the $a/c$ domain structure
formed in the STO, we applied an electric field of $2\times10^{5}$\,V/m
using a back gate on the substrate. This field should break the symmetry
between $a$ and $c$ domains and lead to a different domain structure
compared to a zero field case. No difference in resistivity was observed
with and without the electric field.

Our observation of resonant phonon-carrier coupling illuminates a key feature of conduction in LSMO. The coupling manifests strongly near
the $x=0.5$ composition, while films of similar thickness at the
$x=0.2$ composition \citep{CarlosPRLPaper} showed a resistivity
cusp much smaller in magnitude. We relate this effect to an increased
carrier coupling to the $\Gamma_{25}$ phonon in the AF-M phase of
the LSMO. In this phase, the Mn $d_{x^{2}-y^{2}}$ $e_{g}$ orbital
is occupied while the $d_{3z^{2}-r^{2}}$ orbital is depopulated \citep{RefWorks:489}.
This causes the carriers' wavefunctions to be concentrated on the
$xy$ MnO$_{2}$ planes, which underpins the 2D character of metallicity
and ferromagnetism in this phase, in contrast to the 3D character
of the $x=0.2$ to $0.4$ composition range. The transport measurements of our thin films probe carrier hopping in the $x$ and $y$ directions. Perturbation of
the bridging oxygen positions due to the $\Gamma_{25}$ phonon will 
have a larger scattering effect on carriers in the AF-M phase compared to the 3D FM phase. This is because in the AF-M case, the electron density case is concentrated closer to the perturbed oxygens in the $xy$ plane through which the conduction occurs. This
configuration also explains why an out-of-plane magnetic field reduces the effect
of the phonon coupling, as can be seen in Fig. \ref{fig:fits}. The
magnetic field causes the Mn spins to cant so that they are partially
aligned out-of-plane. This allows for some inter-plane hopping and
reduces the confinement of carriers to the $xy$ MnO$_{2}$ planes,
similarly to the ``spin valve'' effect in A-type Nd$_{0.45}$Sr$_{0.55}$MnO$_{3}$
\citep{spinvalve}.

In conclusion, we show how a single phonon mode originating in the substrate extends resonantly
across an epitaxial interface and into the film. The effects of this coupling are amplified by the properties
of both materials: phonon softening in the substrate causes the phonon amplitude
to diverge, while the LSMO's electronic phase and charge distribution are tuned using strain and a ferroelectric gate.

This work was supported by the National Science Foundation under Contract
MRSEC No. DMR-0520495, DMR-1006265, and FENA. Computational resources were provided by Yale
High Performance Computing, partially funded by grant CNS 08-21132
and by TeraGrid/NCSA under grant number TG-MCA08X007.

\bibliographystyle{apsrev}
\bibliography{cusp}

\end{document}